\documentclass[english]{sig-alternate}
\usepackage[T1]{fontenc}
\usepackage[latin9]{inputenc}
\usepackage{algorithm2e}
\usepackage{amsmath}
\usepackage{amssymb}
\usepackage{graphicx}

\makeatletter

\providecommand{\tabularnewline}{\\}

\usepackage{url}
\usepackage{tikz}
\usepackage{relsize}
\usepackage{pgfplots}
\usetikzlibrary{calc}
\usetikzlibrary{snakes}
\usetikzlibrary{decorations.pathreplacing}
\usepackage{seqsplit}

\@ifundefined{showcaptionsetup}{}{%
 \PassOptionsToPackage{caption=false}{subfig}}
\usepackage{subfig}
\makeatother

\usepackage{babel}
\begin{document}
\numberofauthors{2}
\author{
\alignauthor
Tom\'{a}\v{s} Pevn\'{y}\\
       \affaddr{CTU in Prague}\\
       \affaddr{Cisco R\&D Center in Prague}\\
       \affaddr{Prague, Czech Republic}\\
       \email{pevnak@gmail.com}
\alignauthor
  Petr Somol\\
    \affaddr{Cisco R\&D Center in Prague}\\
    \affaddr{UTIA, Czech Academy of Sciences}\\
    \affaddr{Prague, Czech Republic}\\
    \email{psomol@cisco.com}
}

\title{Discriminative models for multi-instance problems with tree-structure}
\maketitle
\begin{abstract}
Modeling network traffic is gaining importance in order to counter
modern threats of ever increasing sophistication. It is though surprisingly
difficult and costly to construct reliable classifiers on top of telemetry
data due to the variety and complexity of signals that no human can
manage to interpret in full. Obtaining training data with sufficiently
large and variable body of labels can thus be seen as prohibitive
problem. The goal of this work is to detect infected computers by
observing their HTTP(S) traffic collected from network sensors, which
are typically proxy servers or network firewalls, while relying on
only minimal human input in model training phase. We propose a discriminative
model that makes decisions based on all computer's traffic observed
during predefined time window (5 minutes in our case). The model is
trained on collected traffic samples over equally sized time window
per large number of computers, where the only labels needed are human
verdicts about the computer as a whole (presumed infected vs. presumed
clean). As part of training the model itself recognizes discriminative
patterns in traffic targeted to individual servers and constructs
the final high-level classifier on top of them. We show the classifier
to perform with very high precision, while the learned traffic patterns
can be interpreted as Indicators of Compromise. In the following we
implement the discriminative model as a neural network with special
structure reflecting two stacked multi-instance problems. The main
advantages of the proposed configuration include not only improved
accuracy and ability to learn from gross labels, but also automatic
learning of server types (together with their detectors) which are
typically visited by infected computers.
\end{abstract}

\section{Motivation}

\begin{figure*}
\begin{centering}
\includegraphics[width=1\textwidth]{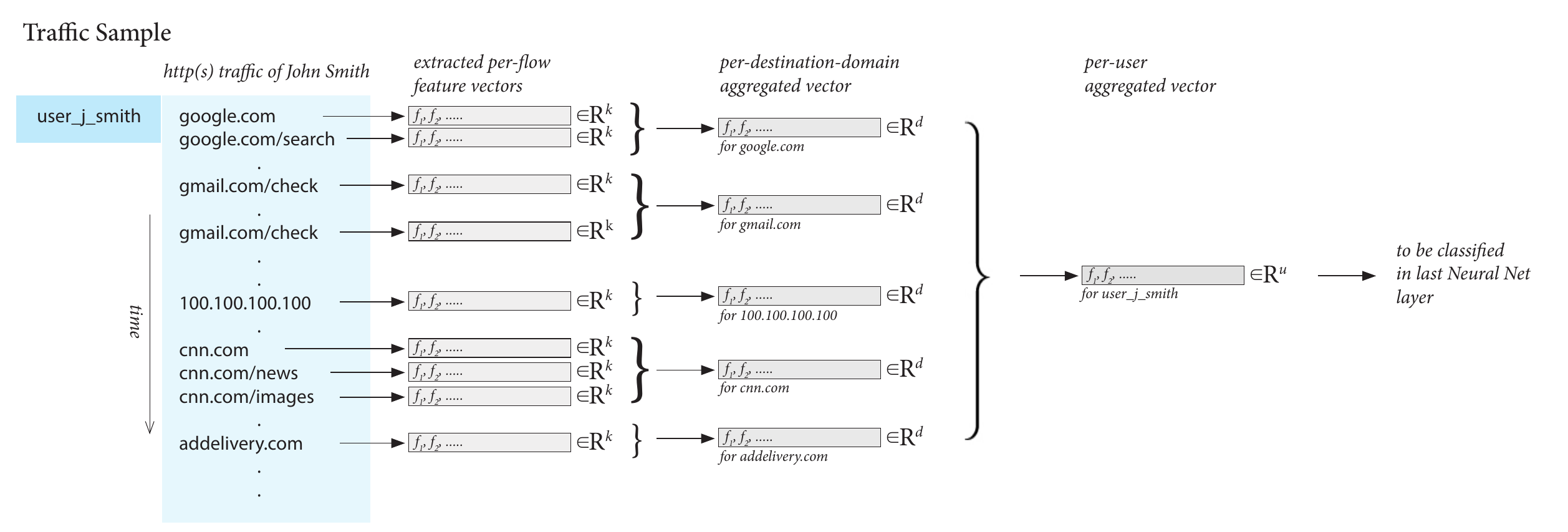}
\par\end{centering}
\caption{\label{fig:Motivation}Sketch of the traffic of a single computer.}
\end{figure*}
\label{sec:Motivation}In network security it is increasingly more
difficult to react to influx of new malicious programs like trojans,
viruses and others (further called malware). Traditional defense solutions
rely on identifying pre-specified patterns (called signatures) known
to distinguish malware in incoming network connections, e-mails, locally
stored programs, etc. But signature-matching now looses breath with
the rapid increase in malware sophistication. Contemporary sophisticated
malware deploys many evasion techniques such as polymorphism, encryption,
obfuscation, randomization, etc, which critically decrease recall
of signature-based methods. The perpendicular approach is identifying
infected computers on the basis of their behavior, i.e., usually by
monitoring and evaluating network activity or system calls. The advantage
of the latter approach is higher recall, because it is much harder
to evade behavior-based detection. E.g., computers infected by spamming
malware almost inevitably display an increase in number of sent e-mails.
Click-fraud, where infected computers earn money to the originator
of infection by showing or accessing advertisements, is another example
where the increased volume of certain traffic is a good indicator
of compromise. On the other hand, behavior-based malware detection
frequently suffers from higher false positive rate compared to signature
based solutions.

Machine learning methods have been recently in focus due to their
promise to improve false-positive-rate of behavioral malware detection\cite{alpcan2010network}.
However, the use of off-the-shelf machine learning methods to detect
malware is typically hindered by the difficulty to obtaining accurate
labels, especially if classification is to be done on the level of
individual network connections (TCP flow, HTTP request, etc.)\cite{Mahoney:2002:LNM:775047.775102,4738466}.
Even for an experienced security analyst it is almost impossible to
determine which network connections are initiated by malware and which
by a benign user or application,\footnote{Even though one has access to the machine infected by malware and
obtain hashes of processes issuing the connection, malicious browser-plugins
will have hash of the browser, which is a legitimate application and
this renders this technique useless. The database of hashes used to
identify malware processes might not be complete yielding to incomplete
labeling.} since malware often mimics behavior of benign connections. We have
observed malware connecting to \texttt{google.com} for seemingly benign
connection checks, displaying advertisements, or sending e-mail as
mentioned above. Labeling individual network connections is thus prohibitive
not only due to their huge numbers but also due to ambiguity of individual
connection's classification. Automatic and large-scale training of
accurate classifiers is thus very difficult. 

In this work we sidestep this problem by moving the object of classification
one level up, i.e., instead of classifying individual connections
we classify the computer (a collection of all its traffic) as a whole.
The immediate benefit is twofold. First, the labeling is much simpler,
as it is sufficient to say ``this computer is infected / clean''
rather than ``this connection has been caused by malware''. Second,
a grouping of connections provides less ambiguous evidence than a
single connection (see cases described above where a single access
of ad server does not tell much, but multitude of such accesses does).
This latter property is in fact the main motivation behind our present
work.

The biggest obstacle in implementing a classifier on basis of all
observed traffic is the variability in the number of network connections
(hereafter called \emph{flows}). This property effectively rules out
majority of machine learning algorithms requiring each sample to be
described by a fixed dimensional vector, because the number of observed
flows supposed to characterize one computer can range from dozens
to millions while the flows may vary in information content. Our problem
thus belongs to the family of multi-instance learning (MIL) problems~\cite{Amores2013,foulds2010review}
where one sample is commonly called a \emph{bag} (in our case representing
a computer) and consists of a variable number of \emph{instances}
(in our case one instance is one flow), each described by a fixed
dimensional vector. 

The solution proposed below differs from common MIL paradigm by taking
a step further and representing data not as a collection of bags,
but as a hierarchy of bags. We show that such approach is highly advantageous
as it effectively utilizes natural hierarchy inherent to our data.
Flows emitted or observed by one computer can be easily grouped according
to servers they connect to (these groups are called \emph{sub-bags)},
so that the \emph{bag} representing the particular computer becomes
a collection of \emph{sub-bags}. This hierarchy can be viewed as a
tree with leafs representing flows (instances), inner nodes representing
servers (sub-bags), and finally the root representing the computer
(bag). The structure of the problem is shown in Figure~\ref{fig:Motivation}.
Note that trees representing different computers will have different
number of inner nodes and leafs. The proposed classifier exploits
this structure by first modeling servers (sub-bags) on basis of flows
targeted to them and then modeling the computer on top of the server
models. This approach can be viewed as two MIL problems stacked one
on top of the other. In Section~\ref{sec:ProposedSolution} we show
how the hierarchical MIL problem can be mapped into neural-network
architecture, enabling direct use of standard back-propagation as
well as many recent developments in the field of deep learning. Once
trained, the architecture can be used for classification but it can
also be dismantled to identify types of traffic significant for distinguishing
benign from infected computer, i.e., it allows to extract learned
indicators of compromise (IoCs). Finally, using approach similar to
URCA~\cite{5462151}, it is possible to identify particular connections
which made the neural network decide that the computer is infected;
hence effectively providing an explanation of the learned IoC.

Section~\ref{sec:Experiments} demonstrates the proposed approach
on large scale real problem of detecting infected computers from proxy
logs. It is shown that the network can learn to identify infected
computers in protected network, as well as provide sound explanation
of its verdicts to the consumer. Neurons in lower layer are shown
to have learned weak indicators of compromise typical for malware.

The proposed neural architecture is shown to have multiple advantageous
properties. Its hierarchal MIL nature dramatically reduces the cost
of label acquisition. By using labels on high-level entities such
as computers or other network devices the creation of training data
is much simpler. The ability to dismantle the encoded structure is
no less important as it provides definition of learned indicators
of compromise. Finally, it allows human understandable explanation
of classifier verdict as security incident, which simplifies the job
of the network administrator.

This paper is organized as follows. The next section formulates the
problem of multiple instance learning and reviews important work we
build upon. The proposed approach is presented in Section~\ref{sec:ProposedSolution}
Experimental evaluation is provided in Section~\ref{sec:Experiments}. 

\section{Related work}

\label{Sec:RelatedWork} In the following we review the evolution
of paradigms leading to the solution proposed in next chapter.

\subsection{Multi instance learning problem}

\label{subsec:MIL}The pioneering work~\cite{dietterich1997solving}
coined \emph{multiple-instance }or\emph{ multi-instance} learning
as a problem, where each sample $b$ (to be denoted \emph{bag} in
the following) consists of a set of instances $x$, i.e., $b=\{x_{i}\in\mathcal{X}|i\in\{1,\ldots|b|\}\}.$
Each instance $x$ can be attributed a label $y_{x}\in\{-1,+1\},$
but these instance-level labels are not assumed to be known even in
the training set. The sample $b$ was deemed positive, if at least
one of its instances had a positive label, i.e., label of a sample
$b$ is $y=\max_{x\in b}y_{x}.$ For this scenario the prevalent approach
is the so-called \emph{instance-space paradigm}, i.e., to train a
classifier on the level of individual instances $f:\mathcal{X}\mapsto\{-1,+1\}$
and then infer the label of the bag $b$ as $\max_{x\in b}f(x).$ 

\subsubsection{Embedded-Space Paradigm}

Later works (see reviews \cite{Amores2013,foulds2010review}) have
introduced different assumptions on relationships between the labels
on the instance level and labels of bags or even dropped the notion
of instance-level labels and considered only labels on the level of
bags, i.e., it is assumed that each bag $b$ has a corresponding label
$y\in\mathcal{Y},$ which for simplicity we will assume to be binary,
i.e., $\mathcal{{Y}}=\{-1,+1\}$ in the following. The common approach
of the latter type is either to follow a \emph{bag-space paradigm}
and define a measure of distance (or kernel) between bags or to follow
an \emph{embedded-space paradigm} and define a transformation of the
bag to a fixed-size vector.

Since the solution presented in Section~\ref{sec:ProposedSolution}
belongs to the embedded-space paradigm, we describe this class of
methods in necessary detail and adopt the formalism of~\cite{Pevny2016},
which is for our solution essential. The formalism of~\cite{Pevny2016}
is intended for a general formulation of MIL problems, where labels
are assumed only on the level of bags without any labels on the level
of instances. Each bag $b$ consists of a set of instances, which
are viewed as a realization of some probability distribution $p_{b}$
defined over the instance space $\mathcal{X}$. To allow more flexibility
between bags even within the same class, the formalism assumes that
probability distributions $p_{b}$ of different bags are different,
which is captured as $p_{b}$ being realization of a probability $P(p_{b},y$),
where $y\in\mathcal{Y}$ is the bag label.

During the learning process each concrete bag $b$ is thus viewed
as a realization of unknown probability distribution $p_{b}$ that
can be inferred only from groups of instances $\{x\in b|x\sim p_{b}\}$
observed in data. The goal is to learn a discrimination function $f:\mathcal{B}\mapsto\mathcal{Y},$
where \emph{$\mathcal{B}$} is the set of all possible realizations
of all distributions $p\in\mathcal{P}^{\mathcal{X}}$, i.e., $\mathcal{B}=\left\{ x_{i}|p\in\mathcal{P}^{\mathcal{X}},x_{i}\sim p,i\in\{1,\ldots l\},l\in\mathbb{N}\right\} $.
Note that this definition also includes that used in~\cite{dietterich1997solving}.\footnote{Ref.~\cite{dietterich1997solving} assumed labels on instances and
a bag was classified as positive if it contained at least one positive
instance. In the used general formulation this corresponds to the
case, where in each positive bag exist instances that never occur
in negative bags, which means that the difference of support of positive
and negative probability distributions is non-empty, i.e., $p_{+}\backslash p_{-}\neq\textrm{Ø},$
where $p_{+}\sim P(p|+)$ and $p_{-}\sim P(p|-).$} 

Methods from \emph{embedded space-paradigm}~\cite{Amores2013,foulds2010review}
first represent each bag $b$ as a fixed-dimensional vector and then
use any machine learning algorithm with samples of fixed dimension.
Therefore the most important component in which most methods differ
is the embedding. Embedding of bag $b$ can be generally written as

\begin{equation}
\left(\phi_{1}(b),\phi_{2}(b),\ldots,\phi_{m}(b)\right)\in\mathbb{R}^{m}.\label{eq:embedding-1}
\end{equation}
with individual projection $\phi_{i}:\mathcal{B}\mapsto\mathbb{R}$
being 
\begin{equation}
\phi_{i}=g\left(\left\{ k(x,\theta_{i})\right\} _{x\in b}\right),\label{eq:projection-phi}
\end{equation}
where $k:\mathcal{X}\times\Theta\mapsto\mathbb{R}_{0}^{+}$ is a suitably
chosen distance function parametrized by parameters $\theta$ (also
called dictionary items) and $g:\cup_{n=1}^{\infty}\mathbb{R}^{k}\mapsto\mathbb{R}$
is the pooling function (e.g. minimum, mean or maximum). Most methods
differ in the choice of aggregation function $g,$ distance function
$k,$ and finally in selection of dictionary items $\theta\in\Theta$. 

\subsection{Simultaneous Optimization of Embedding and Classifier}

\begin{figure}[t]
\begin{centering}
\includegraphics[scale=0.8]{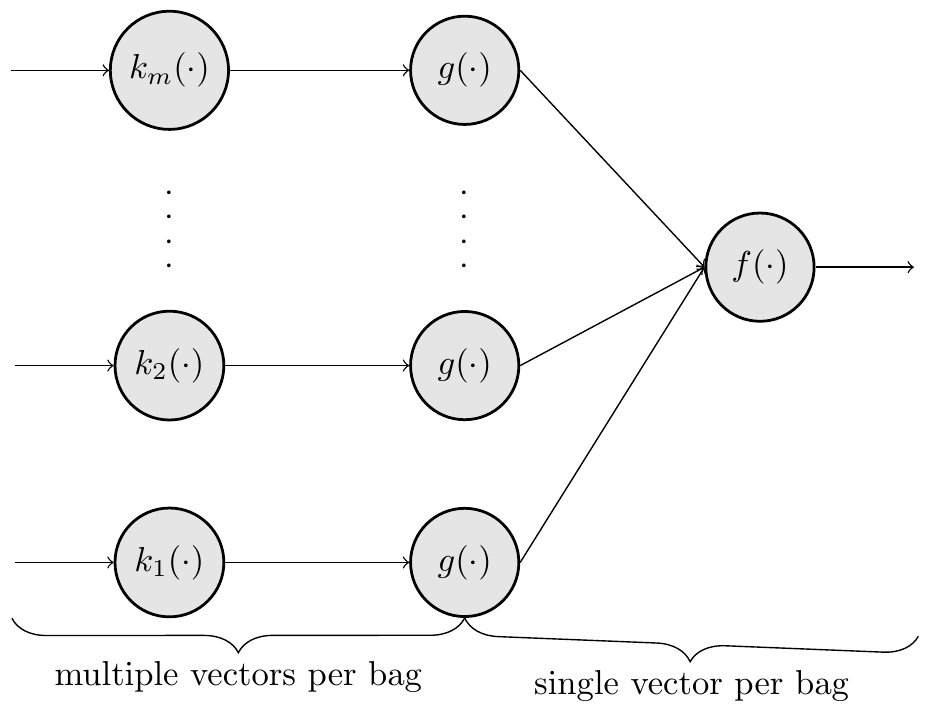}
\par\end{centering}
\caption{\label{fig:simplemil}Neural network optimizing the embedding in embedding-space
paradigm.}
\end{figure}
 The important novelty introduced in~\cite{Pevny2016} is that embedding
functions $\{\phi_{i}\}_{i=1}^{m}$ are optimized simultaneously with
the classifier that uses them, as opposed to the prior art where the
two optimization problems are treated indepedently. Simultaneous optimization
is achieved by using the formalism of neural network, where one (or
more) lower layers followed by a pooling layer implement the embedding
function $\phi,$ and subsequent layers implement the classifier that
is thus built on top of bag representation in form of a feature vector
of fixed length. The model is sketched in Figure~\ref{fig:simplemil}
with a single output neuron implementing a linear classifier once
the embedding to a fixed-length feature representation is realized.
The neural network formalism enables to optimize individual components
of the embedding function as follows.
\begin{itemize}
\item Lower layers (denoted in Figure~\ref{fig:simplemil} as $\{k_{i}\}_{i=1}^{m})$
before pooling identifies parts of the instance-space $\mathcal{X},$
where the probability distributions generating instances in positive
and negative bags differs the most with respect to the chosen pooling
operator.
\item The pooling function $g$ can be either fixed to mean or maximum,
or other pooling function such that it is possible to calculate gradient
with respect to its inputs. The pooling function itself can have parameters
that can be optimized during learning, as was shown e.g. in~\cite{gulcehre2014learned},
where the pooling function has form $\sqrt[\leftroot{-1}\uproot{3}q]{\frac{1}{|b|}\sum_{i\in b}|x_{i}|^{q}}$
with the parameter $q$ being optimized.
\item Layers after the pooling (denoted in Figure~\ref{fig:simplemil}
as $f(\cdot)$) optimize the classifier that already uses representation
of the bag as vector of fixed dimension. 
\end{itemize}
The above model is very general and allows automatic optimization
of all parameters by means of back-propagation, though the user still
needs to select the number of layers, number of neurons in each layer,
their transfer function, and possibly also the pooling function.

\section{The proposed solution}

\label{sec:ProposedSolution} In the light of the previous paragraph,
the problem of identifying infected computers can be viewed as two
MIL problems, one stacked on top of the other, where the traffic of
a computer $b$ is generated by a two-level generative model.

\subsection{Generative Model}

Let us denote $\mathcal{S}$ the set of all servers accessible by
any computer. Let $\mathcal{S}_{c}\subseteq\mathcal{S}$ denote the
selection of all servers accessed from computer $c$ in given time
frame. The communication of computer $c$ with each server $s\in\mathcal{S}_{c}$
consists of a group of flows $x\in\mathcal{X}$ that are viewed as
instances forming a \emph{first-level bag} $b_{s}$. Bag of flows
$b_{s}$ is thus viewed as a realization of some probability distribution
$p_{b_{s}}\in\mathcal{P}^{\mathcal{X}}$ . 

We imagine that every server $s$ is associated with a type $t(s),$
which influences the probability distribution of the flows $p_{b_{s}}$.
Accordingly, each first-level bag $b_{s}$ is realized according to
$p_{b_{s}}$ which itself is a realization of a probability distribution
$P(p_{b_{s}},t(s)).$ This captures the real-world phenomenon of user's
interaction with some server (e.g., e-mail server) being different
from that of a different user communicating with the same server,
as well as the fact that different types of servers impose different
communication patterns.

In view of the above we can now consider computer $c$ to be the \emph{second-level
bag} consisting of a group of first-level bags $b_{s}$. Similarly
to the above we assume $c$to be a realization of probability distribution
$p_{c}\in\mathcal{P}^{\mathcal{B}}$ where \emph{$\mathcal{B}$} is
the set of all possible realizations of all distributions $p\in\mathcal{P}^{\mathcal{X}}$.
Probability distribution $p_{c}$ is expected to be different for
each computer, particularly we assume this to be true between infected
and clean computers labeled by $y\in\{-1,+1\}$. Probability distribution
$p_{c}$ is thus viewed as realization of a probability distribution
$P(p_{c},y)$. This captures the real-world observation that infected
computers exhibit differences in communication patterns to servers,
both in selection of servers and inside individual connections to
the same server.

\begin{algorithm}
\SetKwInOut{Output}{output}
\SetKwInOut{Input}{input}
 \Input{$y\in\{-1,+1\}$ label marking computer  as clear or infected}
 \Output{Set of flows $\mathcal{F}$ of one computer}
 1. sample a distribution $p_{c}$ of servers from $P(p_{c},y)$\;
 2. sample a set of servers $\mathcal{S}_{c}$ from $p_{c}$\;
 3. $\mathcal{F}=\textrm{Ø}$\;
 \ForEach(\%iterate over selected){$s\in\mathcal{S}_{c}$}{
 4. sample distribution $p_{b_{s}}$ of flows from $P(p_{b_{s}},t(s))$\;
  5. sample flows $x$ from $p_{b_{s}}$\;
  6. add sampled flows to all flows, $\mathcal{F}=\mathcal{F}\cup x$\;
 }\
 \caption{\label{alg:model}Generative model of the flows of one computer.}
\end{algorithm}

The model imposes a generative process as illustrated in Algorithm~\ref{alg:model}. 

The proposed multi-level generative model opens up possibilities to
model patterns on the level of individual connections to server as
well as on the level of multiple servers' usage. In the following
we discuss the implementation and show the practical advantages on
large-scale experiments.

\subsection{Discriminative model}

\label{subsec:DiscriminativeModel}
\begin{figure*}
\begin{centering}
\includegraphics[clip,width=0.9\textwidth]{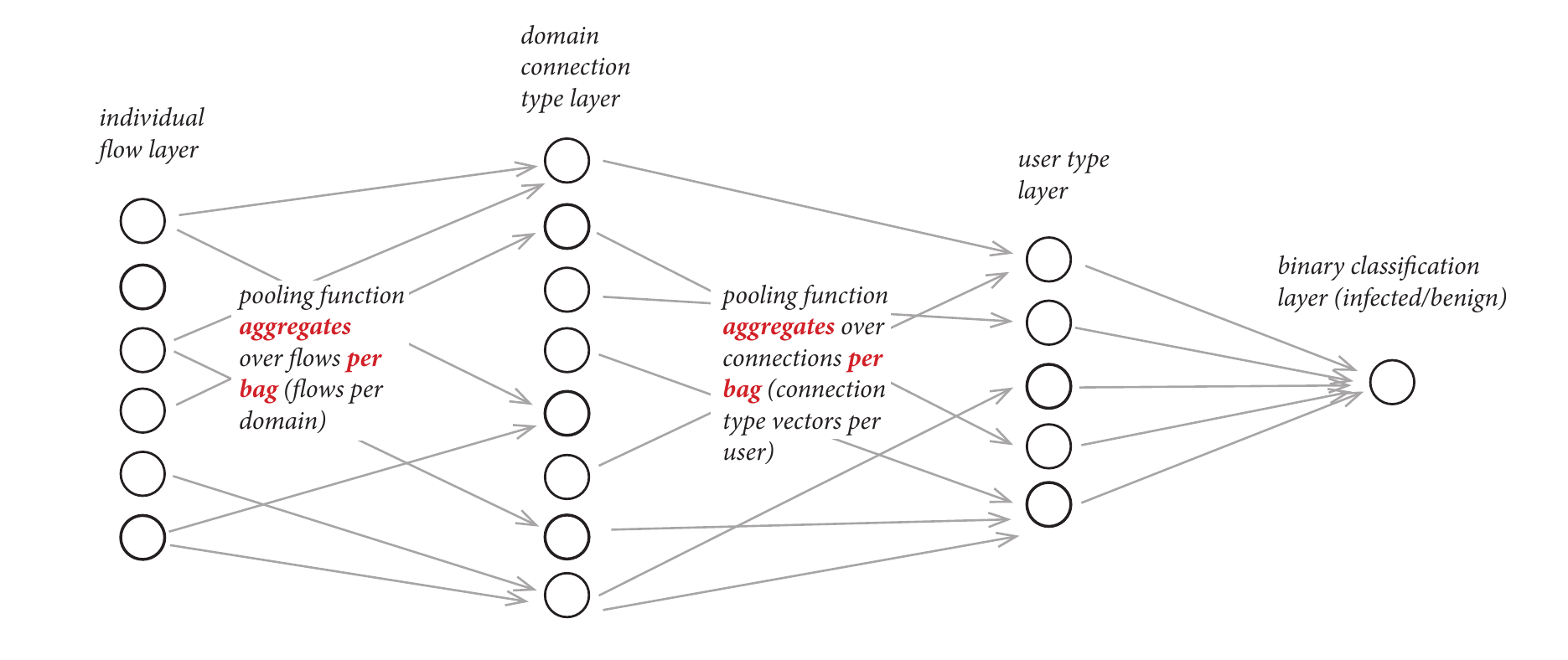}
\par\end{centering}
\caption{\label{fig:MultiMIL}Hierarchical MIL}
\end{figure*}
 The rationale behind the discriminative model closely follows the
above generative model by breaking the problem into two parts: classifying
the computer on basis of types of contacted servers and classifying
type of the server on basis of flows exchanged between the server
and the client. 

Let's assume that each contacted server is described by a feature
vector of fixed dimension, which can be as simple as one-hot encoding
of its type $t(s).$ Then the problem of classifying the computer
becomes a MIL problem with bag being the computer and instances being
servers. The problem is of course that type servers $t(s)$ are generally
unknown and we cannot imagine to manually create a mapping between
server IP or domain name and server type. To make the problem even
more difficult, the same server can be used differently by different
computers, and therefore it can be of different type for each of them.
One can indeed learn a classifier that would predict the server type
from flows between the computer and the server, which again corresponds
to MIL classifier with the bag being the server and instances being
the flows, but the problem of labeled samples for training the classifier
is non-trivial and it is unlikely that we will have known all types
of servers. Moreover, since we are learning a discriminative model,
we are interested in types of server occurring with different probabilities
in clean and infected computers.

To side step this problem we propose to stack MIL classifier on the
level of computers on top of the MIL classifier on the level of servers.
Since both MIL classifiers are realized by a neural network described
in the previous chapter, we obtain one (bigger) neural network with
all parameters optimizable using standard back-propagation and importantly
using labels only on the level of bag (computer). This effectively
removes the need to know types of servers $t(s)$ or learn classifier
for them, because the network learns that automatically from the labels
on the level of computers. The caveat is that the network learns only
types of servers occurring with different probabilities in clean and
infected computers. 

The idea in its simplest incarnation is outlined in Figure~\ref{fig:MultiMIL}.
The distinctive feature is the presence of two pooling layers reflecting
two MIL problems dividing the network into three parts. The first
part part up to the first pooling included implements embedding of
sub-bags into a finite-dimensional vector (modeling servers on basis
of flows). After the first pooling each sub-bag (server) is represented
by one finite-dimensional vector. Similarly the second part starting
between the first pooling up to the second pooling included embeds
sub-bags into a finite dimensional vector characterizing each bag
(computer). Finally, the third part starting after to second pooling
implements the final classifier. 

The right choice of the pooling function is not straightforward with
many aspects to be taken into the consideration.
\begin{itemize}
\item $\mathrm{Mean}$ function should be theoretically better~\cite{muandet2012learning},
since it is more general. The advantage of $\mathrm{mean}$ pooling
function has been experimental demonstrated in~\cite{Pevny2016}.
\item If malware performs few very distinct types of connections (e.g. connection
checks) even though they would go to well known servers, $\max$ functions
can identify them whereas $\mathrm{mean}$ function might suppress
them among the clutter caused by many connections of legitimate applications.
This problem has been recently studied in~\cite{boureau2010theoretical}
in context of natural images.
\item The number of contacted servers and flows to servers varies between
computers and $\mathrm{max}$ pooling is more stable then $\mathrm{mean}.$
\item The training with $\max$ pooling is approximately six times faster,
since the back-propagation is non-zero only for one element entering
the pooling operation (one flow per server and neuron, one server
per computer and neuron).
\end{itemize}

\subsection{Extracting indicators of compromise}

The presented model is based on the assumption that there exist types
of servers contacted with different probability by infected and clean
computers, though one generally does not know much about them. If
these types would not exists, then the probability distributions $p_{c}$
of infected and clean computers would be the same and it would be
impossible to create a reliable detector for them. But if the neural
network has learned to recognize them, items of vector representation
of servers (output of network's first part (from the input to the
first pooling included in Figure~\ref{fig:MultiMIL}) has to have
different probability distributions for clean and infected computers. 

Since the above line of reasoning can be extended to the output of
the layer just before the first pooling function, output of each neuron
of this layer can be viewed as an indicators of compromise, since
it has to contribute to the identification of infected computers.
From a close inspection of flows on which these neurons provide the
highest output a skilled network analyst can figure out, what kind
of traffic it is (a concrete examples are shown in Section~\ref{subsec:IOC}).
Admittedly, these learned IOCs would deliver poor performance if used
alone. But in the neural network they are used together with IOCs
from different servers, which provides a context leading to good accuracy.
Also, once a network administrator annotates these neurons, this annotation
can be used to provide more detailed information about the decisions.

\subsection{Explaining the decision}

\label{subsec:explanation}Neural networks have a reputation being
a black-box in the sense that they do not provide any details about
the decision. In the intrusion detection this behavior is undesirable,
since the investigation of the security has to start from the very
beginning. Therefore giving the investigator explanation why the classifier
view the computer as infected is of great help.

The explanation method relies on the assumption that flows caused
by the infection are additive, i.e. the malware does not block user's
flows but adds its own. This means that if the computer was deemed
infected, by removing right flows (instances) the network should flip
its decision. Although finding the smallest number of such flows is
likely an NP complete problem, a greedy approximation inspired by~\cite{5462151}
performs surprisingly well.

The greedy approximation finds in each iteration set of flows going
to same server (subbag), which causes the biggest decrease of classifier's
output when removed from computer's traffic (in our implementation
positive means infected). Iterations are stopped when classifier's
output becomes negative (clean). The set of all removed subbags is
returned as the explanation in the form: ``This computer was found
infected because it has communicated with these domains''. Examples
of flows to these domains might be obviously supplied.

\subsection{Computational complexity}

The computational complexity is important not only for the training,
but also for the deployment as the amount of network traffic that
needs to be processed can be high. For example Cisco's Cognitive Threat
Analytics~\cite{CTA} processes $10^{10}$ HTTP logs per day. The
hierarchical aggregation inside the network decreases substantially
the computational complexity, since after the first pooling, the network
have one vector for server instead to for one vector per flow yielding
to six fold decrease of the data to be processed. Similarly, after
the second pooling the computer is described just by a single vector
instead of set of vectors, which against decreases the complexity.
Compare this to the prior art on solving MIL with Neural Network~\cite{zhou2002neural},
where the pooling is done after the last linear layer just before
the output, which means that all layers of the network process all
flows. The effect on the computational complexity is tremendous. Whereas
our approach takes approximately five seconds per 100 iterations of
the training, the prior art of~\cite{zhou2002neural} takes 1100
seconds, which is 220 times slower.

\section{Experimental evaluation}

\label{sec:Experiments}Albeit the proposed solution si general and
can be utilized to any type of network traffic, it has been evaluated
in the context of detecting infected computers from logs of web proxies
due to availability of large data to us. Besides, proxy logs are nicer
for human's investigation than for example netflow data. The proxy
logs were collected by Cisco's Cognitive Threat Analytics~\cite{CTA}
from 500 large networks during eight days. The days were picked randomly
from the period from November 2015 till February 2016 with the testing
day being 7th March 2016. Since the total number of infected computers
in the dataset from seven training days was small, we have added data
of infected computers from additional 25 days from the period of training
data. 

Since the data were collected in five-minute long time windows, one
bag consists of all web request of one computer during that window.
Computers were identified either by source IP address or by the user
name provided in proxy logs. Subbags contained requests with the same
host part in the HTTP request. 

Computers (bags) were labeled using Cisco's Cognitive Threat Analytics~\cite{CTA}
such that if one computer had at least one request known to be caused
by malware, the computer was considered to be infected in that five-minute
window. If the same computer in different time window did not have
any malware flows, the bag from that time window was considered as
clean. 

The training set contained data from from approximately 20 million
unique computers out of which 172 013 were infected and approximately
850 000 000 flows, out of which 50 000 000 belonged to infected computers.
The testing set contained data of approximately 3 000 000 computers
out of which 3 000 were infected and approximately 120 000 000 flows
with 500 000 flows belonging to the infecting computers.

We are certain that labeling we have used in this experiment is far
from being perfect. While there will be relatively small number number
of infected computers labeled as clean, there will be quite a lot
of computers labeled as clean while being infected. Despite these
issues, we consider this labeling as a ground truth, because the aim
of the experiments is to demonstrate that the proposed solution can
learn from high-level labels and identify weak indicators of compromise.

The experiments were implemented in author's own library, since popular
libraries for neural networks are not designed for MIL problems. They
do not allow to have samples (bags and sub-bags) of different sizes
(number of instances) which makes the encoding of the hierarchical
structure impossible. Therefore evaluated architectures used simple
building blocks: rectified linear units~\cite{glorot2011deep,muandet2012learning},
mean and maximum pooling functions, and ADAM optimization algorithm~\cite{kingma2014adam}.
Unless said otherwise, ADAM was used with default parameters with
the gradient estimated in each iteration from $1000$ legitimate and
$1000$ infected computers (bags) sampled randomly. This size of the
minibatch is higher then is used in most art about deep learning,
however we have find it beneficial most probably because the signal
to be detected is weaker. Contrary to most state of the art, we have
used weighted Hinge loss function $\max\left\{ 0,1-y\cdot w^{y}\cdot f(x)\right\} $
with $w^{\mathrm{+}}$ being the cost of (false negative) missed detection
and $w^{-}$ being the cost of false positive (false alarms). The
rationale behind Hinge loss is that it produces zero gradients if
sample (bag) is classified correctly with sufficient margin. This
means that gradient with respect to all network parameters is zero,
therefore the back-propagation does not need to be performed, which
leads to considerable speed-up. The learning was stopped after ADAM
has performed $3\cdot10^{5}$ iterations.

The performance was measured using precision-recall curve (PR curve)~\cite{ASI:ASI5090060411}
popular in document classification and information retrieval due to
its better properties for highly imbalanced problems, into which intrusion
detection belongs (in the testing data there is approximately one
infected computer per one thousand clean ones). 

\subsection{Network architecture}

\begin{figure}
\begin{centering}
\subfloat[relu-max-relu-max-lin]{\begin{centering}
\includegraphics[width=0.49\columnwidth]{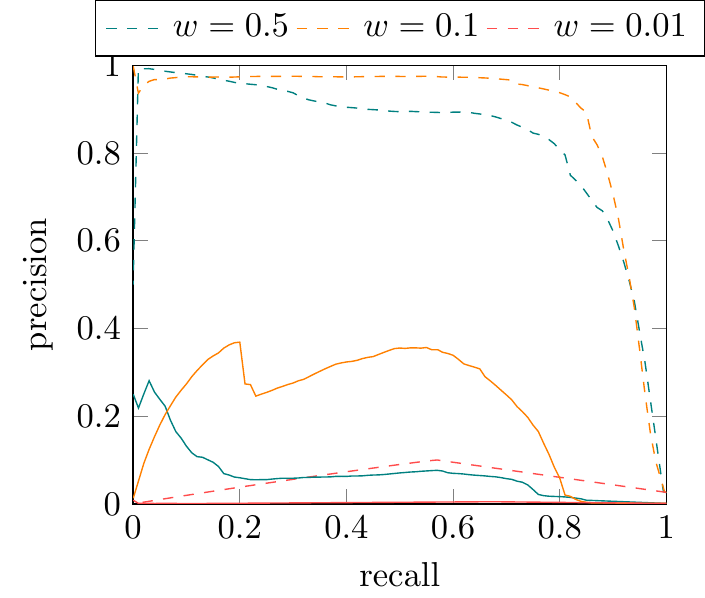}
\par\end{centering}
}\subfloat[relu-mean-relu-mean-lin]{\begin{centering}
\includegraphics[width=0.49\columnwidth]{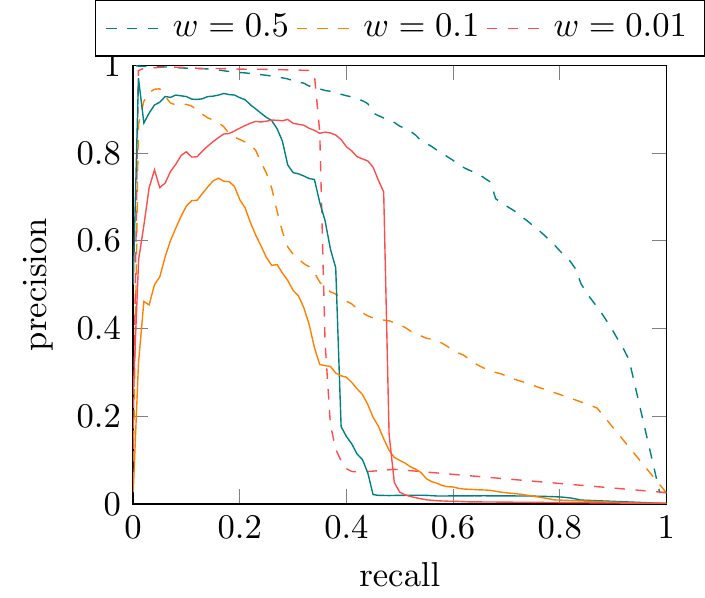}
\par\end{centering}
}\\
\subfloat[relu-max-relu-max-relu-lin]{\begin{centering}
\includegraphics[width=0.49\columnwidth]{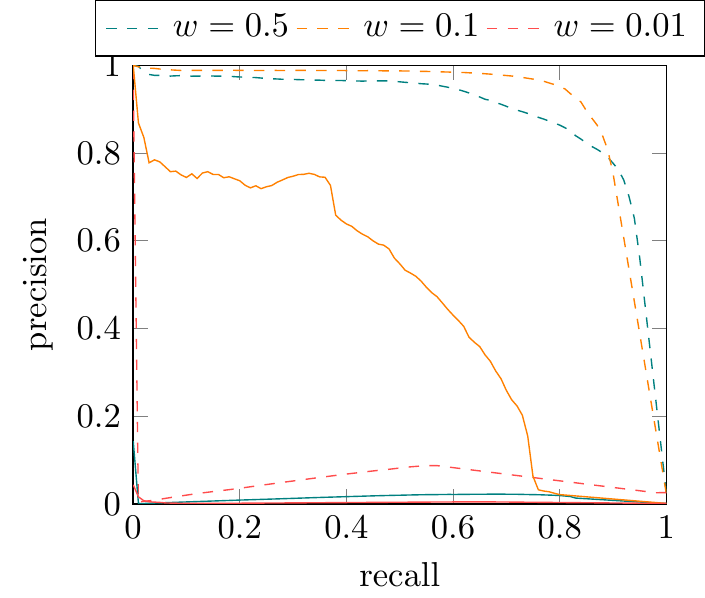}
\par\end{centering}
}\subfloat[relu-mean-relu-mean-relu]{\begin{centering}
\includegraphics[width=0.49\columnwidth]{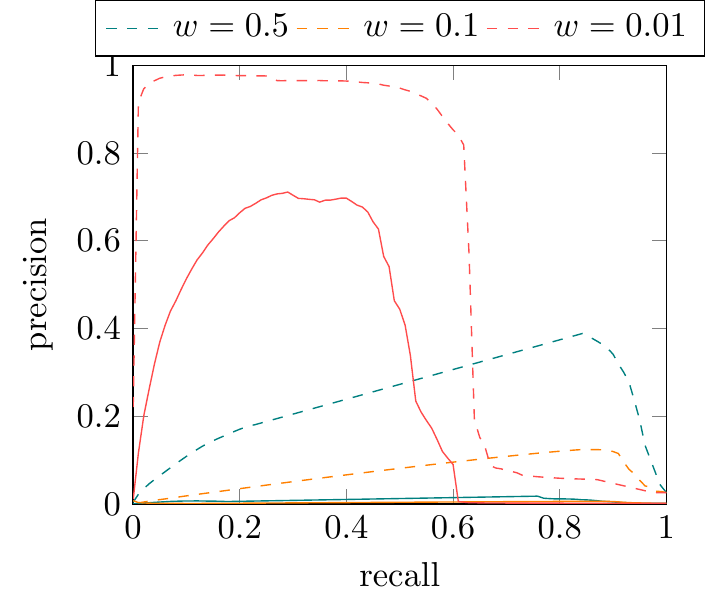}
\par\end{centering}
}\\
\subfloat[relu-max-relu-relu-max-relu-lin]{\begin{centering}
\includegraphics[width=0.49\columnwidth]{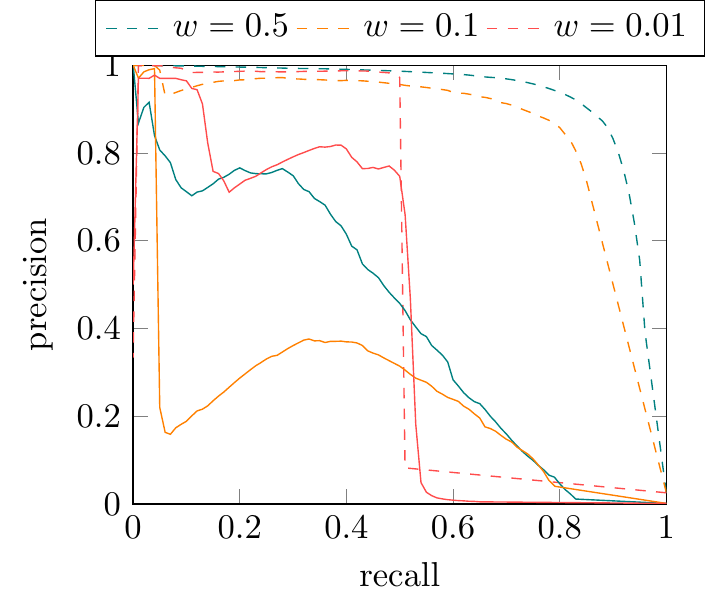}
\par\end{centering}
}\subfloat[\label{fig:relu-mean-relu-relu-mean-relu-li}relu-mean-relu-relu-mean-relu-lin]{\begin{centering}
\includegraphics[width=0.49\columnwidth]{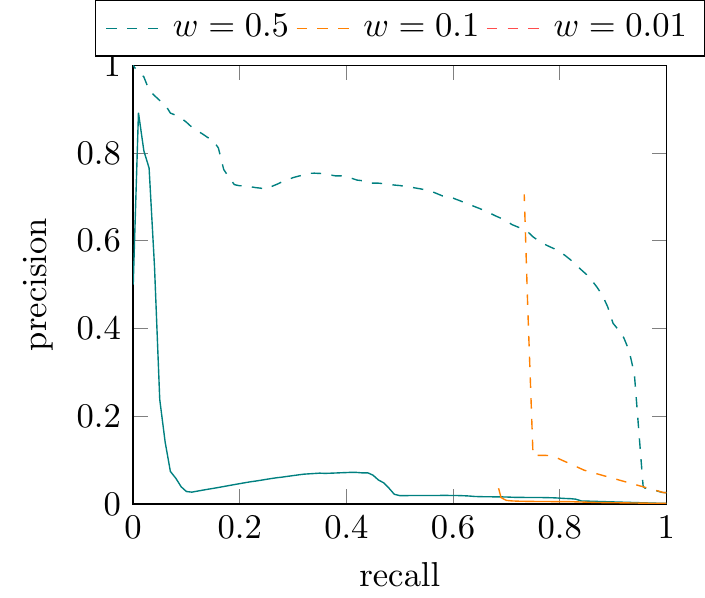}
\par\end{centering}
}
\par\end{centering}
\caption{\label{fig:34features}Precision recall curves of six neural network
architectures utilizing simple $36$ features. Dashed lines shows
the curves estimated on the training set and solid lines shows the
curves estimated from the training set. Networks with PR curves in
the left column used $\mathrm{max}$ pooling function, whereas those
with PR curves in the right column used $\mathrm{mean}$ pooling function.
Captions $w=0.5,$ $w=0.1,$ and $w=0.01$ corresponds to different
costs in weighted hinge loss with cost on false positives (false alarms)
being $w^{-}=1-w$ while that on the false negatives (missed detections)
being $w^{+}=w.$}
\end{figure}
 All evaluated neural networks used simple feature vectors (instances)
with $36$ cheap to compute statistics, such as length of the url,
query and path parts, frequency of vowels and consonants, HTTP status,
port of the client and the server, etc, but not a single feature was
extracted from the hostname. Evaluated neural networks followed the
architecture in Figure~\ref{fig:MultiMIL} with layer of 40 ReLu
neurons before the first pooling, but then differing in: using either
$\mathrm{mean}$ or $\mathrm{max}$ pooling functions; having either
one layer with 40 ReLu neurons or two layers each with 20 ReLu neurons
between first and second pooling; and finally having additional layer
of 20 ReLu neurons  after the second pooling and final linear output
neuron. 

Precision-recall curves of all six evaluated neural networks each
trained with three different costs of errors on false positive (0.9,0.99,0.999)
and false negative (0.1,0.01,0.001) are shown in Figure~\ref{fig:34features}.
On basis of these experiments, we have made following conclusions.
\begin{itemize}
\item Simpler networks with $\mathrm{max}$ pooling function tends to overfit,
as the error on the training set of all three evaluated architectures
is very good (dashed line) but the error on the testing set is considerably
worse. We believe this to be caused by the network to act more like
a complicated signature detector by learning a specific patterns in
flows prevalent in the infected computers in the training set, but
missing in infected computers in testing set.  This hypothesis is
supported by (i) the fact that when we have been creating ground truth,
we have labeled computer as infected if it had at least one connection
known to be caused by malware and (ii) testing data being one month
older then training ones. 
\item Simple networks with $\mathrm{mean}$ pooling with costs of error
$w^{\mathrm{+}}=0.01$ and $w^{-}=0.99$ are amongst the best ones.
Their discrepancy between training and testing error is much lower
then in the case of $\mathrm{max}$ pooling, except the most complicated
architecture~\ref{fig:relu-mean-relu-relu-mean-relu-li}. We believe
this to be caused by the network learning how infected computers behave
(contacting too many advertisement servers) rather than patterns specific
for some type of the malware (like those with $\mathrm{max}$ pooling).
This conclusion is supported by the fact that $\mathrm{max}$ pooling
function can be approximated from the $\mathrm{mean}$ if layers preceding
the aggregation are sufficiently complicated~\cite{7368555}.
\end{itemize}
Interesting feature is sharp drop in precision of certain architectures,
which we attribute to the fact that some infections cannot be detected
with used simple 34 features.

\subsection{Indicators of compromise}

\label{subsec:IOC}Since one of the main features of the proposed
architecture is an ability to learn indicators of compromise IOCs,
below it is shown to what types of traffic neurons in the layer just
before the first pooling are sensitive. The sensitivity was estimated
from infected computers in the testing set for the simplest architectures
(top row in Figure~\ref{fig:34features}) with $\mathrm{mean}$ and
$\mathrm{max}$ pooling functions. 

We have not observed much difference between IOCs learned by network
with $\mathrm{mean}$ and $\mathrm{max}$ pooling functions. Learned
IOCs included:
\begin{itemize}
\item tunneling through url (example shown in appendix due to its length);
\item sinkholed domains such as \texttt{\seqsplit{hxxp://malware.vastglows.com}},
\texttt{\seqsplit{hxxp://malware.9f6qmf0hs.ru/a.htm?u=396923}}, \texttt{\seqsplit{hxxp://malware.ywaauuackqmskc.org/}}.
\item domains with repetitive characters such as \texttt{\seqsplit{hxxp://wwwwwwwwwwwwvwwwwwwwwwwwwwwwwwwvwwwwwwwwwwwwwwwwwwwwwwwwwwwwvww.com/favicon.ico}}
or \texttt{\seqsplit{hxxp://ibuyitttttttttttttttttttttttttttttttttttibuyit.com/xxx.zip}};
\item https traffic to raw domains such as\texttt{ \seqsplit{hxxps://209.126.109.113/}};
\item subdomain generated by an algorithm on a hosting domain, for example
\texttt{d2ebu295n9axq5.webhst.com}, \texttt{\seqsplit{d2e24t2jgcnor2.webhostoid.com}},
or\texttt{ }\seqsplit{dvywjyamdd5wo.webhosteo.com};
\item Download of infected seven-zip: \texttt{\seqsplit{d.7-zip.org/a/7z938.exe}}\footnote{We refer to \url{hxxps://www.herdprotect.com/domain-d.7-zip.org.aspx} for confirmation that this is indeed malware related.}.
\end{itemize}

\subsection{Example of explanation}

\begin{table}
\centering{}%
\begin{tabular}{cl}
NN  & \tabularnewline
output & url\tabularnewline
\hline 
4.84 & \texttt{hxxp://www.inkstuds.org/?feed=podcast}\tabularnewline
2.07 & \texttt{hxxp://feeds.podtrac.com/YxRFN5Smhddj}\tabularnewline
0.21 & \texttt{hxxps://www.youtube-nocookie.com/}\tabularnewline
0.18 & \texttt{hxxps://upload.wikimedia.org/}\tabularnewline
\hline 
\end{tabular}\caption{\label{tab:Explanation}Example output of the explanation of an incident.}
\end{table}
Table~\ref{tab:Explanation} shows an explanation of the simplest
evaluated neural network with maximum pooling functions. The explanation
consists of a list of domains with examples of requests to them as
they have been identified by the greedy algorithm described in Section~\ref{subsec:explanation}.
The column captioned ``NN output'' shows, how the output of the
neural net decreases as flows to individual domains are iteratively
removed. 

At the time of writing this paper, the last three domains were all
involved in the communication with some malware samples according
to VirusTotal~\cite{vt}. Searching further on a web we have found
this article\footnote{\url{http://inkstuds.tumblr.com/post/139553865057/started-my-day-with-the-inkstuds-site-getting}}
stating that \texttt{www.inkstuds.org} have been hacked and used to
serve malware.

\section{Conclusion}

We have introduced stacked Multiple Instance Learning architecture,
where data is viewed not as a collection of bags but as a \emph{hierarchy
of bags}. This extension of MIL paradigm is shown to bring many advantages
particularly for our targeted application of intrusion detection.
The hierarchical model is straightforward to implement, requiring
just a slight modification of a standard neural network architecture.
This enables to exploit vast neural network knowledgebase including
deep learning paradigms. 

The proposed architecture posses key advantages especially important
in network security. First, it requires labels (clean / infected)
only on the high level of computers instead of on single flows, which
dramatically saves time of human analyst constructing the ground truth
and also makes it more precise (it might be sometimes nearly impossible
to determine, if the flow is related to infection or not). Second,
the learned mapping of traffic patterns to neurons can be extracted
to obtain human understandable Indicators of Compromise. Third, it
is possible to identify flows, which have cased the computer to be
classified as infected, which decreases time needed to investigate
the security incident.

The advantages of the proposed architecture were demonstrated in the
context of detecting infected computers from their network traffic
collected on the proxy server. It has been shown that the neural network
can detect infected computers, learn indicators of compromise in lower
layers from high-level labels, and provide sound explanation of the
classification.

\bibliographystyle{plain}

\appendix

\section{Types of learned IOCs}
\begin{itemize}
\item tunneling through urls
\end{itemize}
\texttt{\seqsplit{hxxp://call.api.bidmatic.com/event/click/e54ae5b5435b118ca6539752037be726e1d6ccbd297e8ce191ad1304c2d813e9b0739b9699e4f69b370663ef3476aa3a4e6b15fd4dbe392849711a223e5635d088bad54f4aeee18fcf830b72c2c6588f5a3faf4db8cf39b5aa5b1ee77bb5cd4254f666a6295ec4c47c9eea5cdd612bcdd9541430f58e27d2d5f36700526f94106ad7bfae9409dcc7d6897be9e015724fcd66e5564ab56f4e1be62456237f7567d667a95f3b24ea2ef127b75e5cc353104579b047f09c5e01eab79a57935692e9be881eec56c4030a01b4ffa7bcdc72430ffe1a8b091182851016c299a8b343f1cc015f6cc9b36e109334b04bfef24b15acf0b0cb4bad9bd9523dbffe0e0171e6f180ce475c3fdd701a33c6a144f135e8d651f54ca92a4fa572938bc248471991542aba5e5f380d5b00c7931384d0a726b1a27db83ceb1178e7355e1451a9e8f8ac91c7306aff1f23be85849b51dfa52f8bb52f1be5cdf5497d739a8760c7c7178a811d7e2555e864bbd5b32840e65862aac63c266a0c6dd72468ae975982db1135322d604d43b62c1259f22677d15ee2dbd86fdfefe84807c66999d87cdaaa92edf007466f73ee2bc14a6d5ee708649c5f7caf814e4497826308a508d4ff94eb91d55ca2e44e02e2ff8740ac7f1c16135319c38eba9fd50e397edf8a98afbc2e1bd18e82208c6109f253370ca95d035aac4edf6e8ef51ab891b85e5b2bf6e8ce3480bc4c69ac505ca31397f7133716ba5d8652d716999c4ecac7b787f663ac6fb0b32a6b6fe10eb740397e893cb58b49bc2ed18b10944d5e149c5935e367f43d94d074ab8b2f732d34e194be43f7f940}}

\texttt{\seqsplit{hxxp://s.crbfmcjs.info/dealdo/shoppingjs4?b=Chy9mZaMDhnSpxvUzgvMAw5LzczKyxrHpsu3qIuYmMGXCYuYmIuZqsu1qIuYmIu1q24LmJaLmJaLmJaLmJaLmJaLmJaLmJaLmJaLmJaLmJaLmJaLmJaLmJaLmJaLmJaLmJaLmJaLmJaLmJaLmJaLmJaLmJaLmJaLmJaLmJaLmJaLnunUjtiWjtiWjtiWjtiWjtiWjtiWjtiWjtiWjtiWjtiWjtiWjtiWjtiWjtiWjtiWjtiWjtiWjtiWjtiWjtiWjtiWjtiWjtiWjtiWjtiWjtiWjtvdBIuYmcuYmcuYmcuYmcuYmcuYmfrLEMeLmJbSysuYmg1HDgvTyxrPy2eLmJbZzw0UmI1JBgfZysuYmgeLmJa3lweLmJaLmJaLmJaLmJiLnuqLmKmLmJj0AxrSzsuYmIuZqsuYmLrLEMeLmJbSysuYmg1HDgvTyxrPy2eLmJbZzw0UmI1JBgfZysuYmgeLmJa3lweLmJaLn0mLmJbZB3jPBMjVCM9KAsuYmcu3qYuYmde0lJa1lJiWmtaLmJiLmKmLmJjKB21HAw4LmJiLm0eLmJj3D3CUzgLKywn0AwmUCM8LmJiLmKmLmJj1CMWLmJiLm0eLmJjODhrWjtnbjtjgjtjgD3D3lMrPzgfJDgLJlNjVjtjgBwf0zxjPywXLlwrPzgfJDgLJzsuYrJeYnZeZm190zxPHlwXHlw1HDgvTyxrPy2eTC2vTltiTy2XHC2eTys03lweLmJiLmKmLmJjLBMmLmJiLm0eLmJjvveyTocuYmIuYqYuYmNDUyw1LjtiYjtnbjtiYjtiYjtjdjtiYAxndB21yjtiYjtnbjtiYt0SLm0fKzwyWjtiYjtjdjtiYzYuYmIuZqsu3qIu3rcuYqYuYmMrWu2vZC2LVBKLKjtiYjtnbjtiYmtq2ndaXodKYmdu0odG0mtyLmJiLmKmLmJjezwfSugX5jtiYjtnbjtiYBNjJEwnMExvZjtiYjtjdjtiYzg1UjtiYjtnbjtiYzgLKywn0AwmUCM8LmJiLmKmLmJjMAxjZDfrPBwuLmJiLm0eLmJjMywXZzsuYmIu3rczJBhy9mtq2mtu2ntq4odmYoczXBt0WjMnIptG0oszWyxj0BMvYpwnYyMzTyYzOCMq9mtuWmgiZytnInMfJmJDLmJHJnJjLmwuYyMeWodDHytGMAhjKC3jJpsz2zwHPy2XLpszJAgfUBMvSpwnYyMzTy2nYzhjFmJaWmZe2mZe4ndmZmdaWmdaWjNnZzxq9nczHChb0purLywXiDxqMAxr5Cgu9AszLEhq9x18MDha9BNvSBcz2CJ0MBhrPBwu9mtq2ndaXodKYmdG0oszKB209y3jIzM1JANmUAw5MBYzZzwXMps4Mzg9TCMvMzxjYzxi9Ahr0CcuYntnbjti1mKyLmJuYrND3DY5KAwrHy3rPyY5YBYuYntjgBwf0zxjPywXLlwrPzgfJDgLJzsuYntjgDgv6ys1TyxrLBwf0AwnHlwnSyxnHlweTn2eMCgXPBMS9jMHSAw5RpszWCM9KDwn0CZ0MAw5ZDgDYCd0MAwfNpwnSAwvUDdeWmc4UjMnVB2TPzxntDgf0Dxm9y29VA2LLrw5HyMXLza==}}
\end{document}